\documentclass[twocolumn,showpacs,preprintnumbers,amsmath,amssymb,prl]{revtex4}

\usepackage{graphicx}
\usepackage{dcolumn}
\usepackage{bm}
\usepackage{color}

\begin{document}

\preprint{}

\title{Resonant addressing and manipulation of silicon vacancy qubits in silicon carbide}

\author{D.~Riedel$^{1}$}
\author{F.~Fuchs$^{1}$}
\author{H.~Kraus$^{1}$}
\author{S.~V\"{a}th$^{1}$}
\author{A.~Sperlich$^{1}$}
\author{V.~Dyakonov$^{1,2}$}
\author{A.~A.~Soltamova$^{3}$}
\author{P.~G.~Baranov$^{3}$}
\author{V.~A.~Ilyin$^{4}$}
\author{G.~V.~Astakhov$^{1}$} 
\email[E-mail:~]{astakhov@physik.uni-wuerzburg.de}

\affiliation{$^1$Experimental Physics VI, Julius-Maximilian University of W\"{u}rzburg, 97074 W\"{u}rzburg, Germany \\
$^2$Bavarian Center for Applied Energy Research (ZAE Bayern), 97074 W\"{u}rzburg, Germany\\
$^3$Ioffe Physical-Technical Institute, 194021 St.~Petersburg, Russia\\ 
$^4$Saint Petersburg Electrotechnical University, 197376 St.~Petersburg, Russia}

\begin{abstract}
Several systems in the solid state have been suggested as promising candidates for spin-based quantum information processing. In spite of significant progress during the last decade, there is a search  for new systems with higher potential [D.~DiVincenzo, Nature Mat. \textbf{9}, 468 (2010)]. We report that silicon vacancy defects in  silicon carbide comprise the technological advantages of semiconductor quantum dots and the unique spin properties of the nitrogen-vacancy defects in diamond. Similar to atoms, the silicon vacancy qubits can be controlled under the double radio-optical resonance conditions, allowing for their selective addressing and manipulation. Furthermore,  we reveal their long spin memory using pulsed magnetic resonance technique. All these results make silicon vacancy defects in silicon carbide very attractive for quantum applications.  
\end{abstract}

\date{\today}


\maketitle

The double  radio-optical resonance in atoms \cite{Kastle} constitutes the  basis for a unprecedented level of coherent quantum control. Atomic time standards \cite{AtomicClocks} and multi-qubit quantum logic gates \cite{TrappedIons} are among the most known examples. In the solid state, semiconductor quantum dots (QDs) and the nitrogen-vacancy (NV) defects in diamond, frequently referred to as artificial atoms, are considered as the most promising candidates for quantum information processing \cite{QC-QDs,Diamond}.  Nevertheless, such a high degree of quantum control, as achieved in atoms, has not yet been demonstrated in these systems so far. Therefore, there is a search for quantum systems with even more potential \cite{DiVincenzo}. 

Recently, intrinsic defects in silicon carbide (SiC) have been proposed as eligible candidates for qubits \cite{SiC-defects,SiC-SiliconVacancy}. Indeed, they reveal quantum spin coherence even at room temperature \cite{Nut,RT_Divacancy,RT_SiliconVacancy}.  All of these experiments have been carried out under non-resonant optical excitation where all spins are controlled simultaneously. However, for spin-based information processing it is necessary to perform manipulations of selected spins, while the rest should remain unaffected. This demonstration in SiC is still an outstanding task. 

The selective spin control can be realized using a resonant optical excitation. As a rule, inhomogeneous broadening is much larger than the natural spectral linewidth, and such resonant addressing can be done on single centers only. To avoid this problem, we applied a special procedure to "freeze" silicon vacancy ($\mathrm{V_{Si}}$) defects during their growth, allowing to preserve a high homogeneity inherent to Lely crystals. This is confirmed by the extremely sharp optical resonances in our samples. The spectral width of the $\mathrm{V_{Si}}$ absorption lines is several $\mathrm{\mu eV}$ (ca. 1~GHz), which comparable with that of a single QD or a single NV center in diamond. 

We then demonstrate the selective spin initialization and readout  by tuning the laser wavelength together with the spin manipulation by means of electron spin resonance (ESR). Such a double radio-optical resonance control indicates that  the $\mathrm{V_{Si}}$ defects strongly interact with light and are well decoupled from lattice vibrations. The latter is also confirmed by the observation of a long spin-lattice relaxation time ($T_1 = 0.1$~ms). This can be potentially used for the physical implementation of scalable multi-qubit quantum logic gates in solid state devices \cite{Stoneham}, vector magnetometry with nanometer resolution \cite{NV-Magnetometer} and quantum telecommunications via spin-photon coupling \cite{Atom-Entanglement}. 

\begin{figure}[btp]
\includegraphics[width=.45\textwidth]{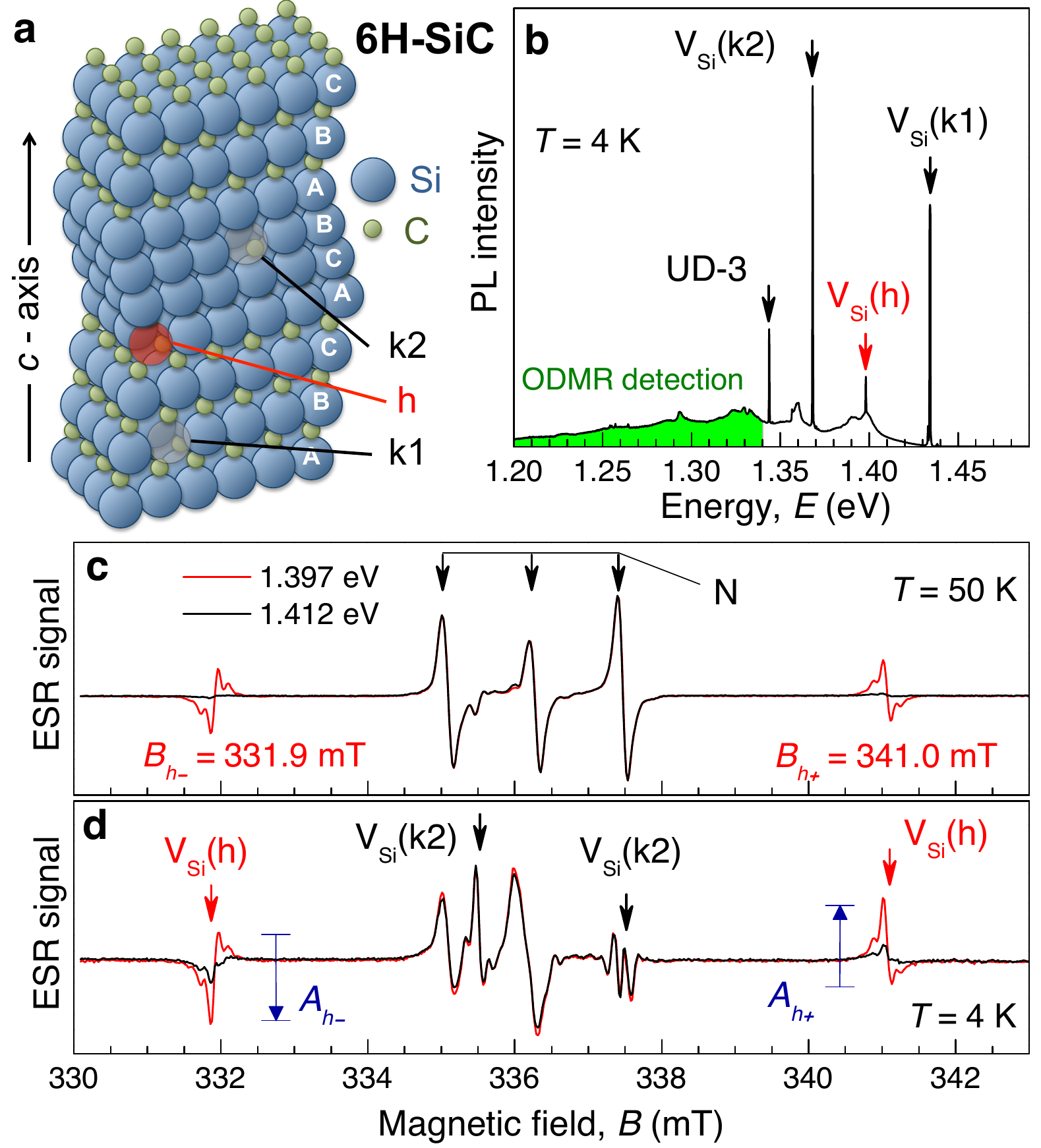}
\caption{(color online). (a) The $\mathrm{V_{Si}}$ defects present at three nonequivalent crystallographic sites of the 6H-SiC lattice, one hexagonal (h) and two quasicubic (k1 and k2).  (b) PL spectrum obtained under excitation with a He-Ne laser ($E = 1.959 \, \mathrm{eV}$). The ZPLs of the corresponding $\mathrm{V_{Si}}$ defects are labeled by arrows. The shaded area from $1.20$ to $1.34 \, \mathrm{eV}$ indicates the spectral range, where  the ODMR of Fig.~\ref{fig4} is detected.  (c,d) Light-enhanced electron spin resonance (ESR) under excitation into the $\mathrm{V_{Si}(h)}$ ZPL ($E_h =1 .397 \, \mathrm{eV}$) and above-ZPL excitation ($E =1.412 \, \mathrm{eV}$), recorded at temperatures $T = 50$ and $4  \, K$. The $\mathrm{V_{Si}(h)}$ ESR lines are observed at magnetic fields $B_{h-} =331.9 \, \mathrm{mT}$ and $B_{h+}  =341.0 \, \mathrm{mT}$. The ESR experiments are performed in a X-band spectrometer ($\nu_{ESR} = 9.43451 \, \mathrm{GHz}$, $B \| c$). } \label{fig1}
\end{figure}

SiC is a wide-gap semiconductor ($E_{g}^{\mathrm{6H-SiC}} = 3.05$~eV) and possesses a spectrum of unique mechanical, electrical and thermophysical properties making it appropriate for many demanding applications \cite{SiC2009}. The material properties of SiC, including intrinsic defects, are being investigated for decades. In particular, the $\mathrm{V_{Si}}$ defects have been identified in the 80's \cite{Ilyin1981}. SiC exists in about 250 crystalline forms, called polytypes, which are variations of the same chemical compound that can be viewed as layers stacked in a certain sequence. Here we concentrate on the polytype 6H-SiC [see Fig.~\ref{fig1}(a)]. It has the stacking sequence ABCACB and is characterized by three nonequivalent crystallographic sites, one hexagonal (h) and two quasicubic (k1, k2). 

The investigated 6H-SiC samples of high crystalline quality have been grown by the modified Lely method. The high-temperature ($2700^{\circ}$C) seedless crystal growth is driven by the temperature gradients within the crucible, resulting in a pressure gradient and thus, in a mass transport. The process is followed by the subsequent fast cooling, which "freezes" the defects within the lattice at low densities (for details see  \cite{Lely}).
Taking into account the retrograde character of the nitrogen solubility with temperature, the doping level of the 6H-SiC crystals is below $10^{17} \, \mathrm{cm^{-3}}$ and the compensation degree of nitrogen donors is ca. 20\%. The $\mathrm{V_{Si}}$ concentration  lies below this value and we estimate that ca. $10 ^{12}$ $\mathrm{V_{Si}}$ are probed in our experiments. 

In all experiments, the sample was mounted in a liquid helium flow cryostat with a temperature controller. PL was excited by a He-Ne laser and detected by a CCD camera linked to a 800-mm monochromator.  For the resonance excitation we used a tunable diode laser system with a linewidth below 1~MHz.  The ESR experiments were performed in a home-modified X-band spectrometer, with the magnetic field applied parallel to the $c$ axis of 6H-SiC. The magnetic field inside the spectrometer was modulated at a frequency of 100~kHz and the ESR detection was locked-in. 

Figure~\ref{fig1}(b) presents a typical low-temperature ($T = 4 \, \mathrm{K}$) photoluminescence (PL) spectrum of our 6H-SiC sample  obtained with sub-bandgap excitation with a HeNe laser ($E = 1.959 \, \mathrm{eV}$). The PL consists of sharp zero phonon lines (ZPLs) and their sideband phonon replicas. Three of these ZPLs -- labeled as $\mathrm{V_{Si}(h)}$, $\mathrm{V_{Si}(k1)}$ and $\mathrm{V_{Si}(k2)}$ -- originate from silicon vacancies at the corresponding crystallographic sites \cite{Wagber2000}. The origin of the UD-3 ZPL is still under debate \cite{UD-3}, and we do not discuss it in what follows. 

We now present light-enhanced electron spin resonance (ESR) experiments. It has been shown that the ground state of silicon vacancies is a high-spin state \cite{GroundS}.  In 6H-SiC with hexagonal crystal structure the $\mathrm{V_{Si}}$ defects have the point group symmetry $\mathrm{C_{3V}}$, allowing for a zero-field spin splitting ($\Delta$). The application of an external magnetic field leads to a further splitting of the $\mathrm{V_{Si}}$ ground state.  When the difference between spin-split sublevels meets the ESR frequency ($\nu_{ESR}$) the electron spin resonance occurs. The energy diagrams describing the optical transitions, the spin pumping scheme and the related RF transitions with respect to the multiplicity of the ground state are discussed in \cite{SupplementalMaterial}. In case of the silicon vacancy, two ESR lines should appear for each $\mathrm{V_{Si}}$ site at magnetic fields $B_-$ and $B_+$ \cite{SiC-SiliconVacancy,SupplementalMaterial}. If the external magnetic field is applied parallel to the $c$ axis of 6H-SiC, there is the following interconnection between $B_{\pm}$,  $\nu_{ESR}$ and $\Delta$
\begin{equation}
h \nu_{ESR}= \mp \Delta + g_e \mu_B B_{\pm} \,.
 \label{zero-field-splitting}
\end{equation}
Here, $g_e = 2.0$ is the electron g-factor and $\mu_B = 5.79 \times 10^{-5} \, \mathrm{eV / T}$ is the Bohr magneton. The amplitudes ($A_{\pm}$) of the $\mathrm{V_{Si}}$ ESR lines depend on the population difference between the particular spin sublevels involved. In the dark  the difference is due to the Boltzmann factor, and the amplitudes $A_{\pm}$ are negligibly small.  The optical excitation of $\mathrm{V_{Si}}$ defects and following relaxation preferentially pump the system into certain spin sublevels of the $\mathrm{V_{Si}}$ ground state.  This results in light-enhanced ESR, as exactly observed in our experiments presented in Figs.~\ref{fig1}(c) and (d).  

First, we discuss the high temperature data [Fig.~\ref{fig1}(c)].  The three lines in the magnetic field range from 335.0 to 337.5~mT are the well known ESR fingerprint of the nitrogen (N) donor in 6H-SiC \cite{SiC2009}, being independent of the optical excitation. The pair of outer lines at magnetic fields $B_{h-} =331.9 \, \mathrm{mT}$ and $B_{h+}  =341.0 \, \mathrm{mT}$ appears in the ESR spectrum under optical excitation into the  $\mathrm{V_{Si}(h)}$ ZPL ($E_h =1 .397 \, \mathrm{eV}$). Using Eq.~(\ref{zero-field-splitting}) we obtain the zero-field spin splitting in the $\mathrm{V_{Si}(h)}$ defect   $\Delta_h = 0.527 \, \mathrm{\mu eV}$ (127~MHz), which is in agreement with the earlier reported value \cite{Bard2000}. 

At cryogenic temperatures, the  $\mathrm{V_{Si}(h)}$ ESR lines are also observed under the above-ZPL excitation ($E =1.412 \, \mathrm{eV}$), as shown in Fig.~\ref{fig1}(d).  In addition, the pair of $\mathrm{V_{Si}(k2)}$ resonances appears.  Again using Eq.~(\ref{zero-field-splitting}) we obtain the zero-field spin splitting in the $\mathrm{V_{Si}(k2)}$ defect  $\Delta_{k2} = 0.11 \, \mathrm{\mu eV}$ (27~MHz) \cite{Bard2000}. 

Remarkably, under the optical excitation the  $B_-$ and $B_+$-lines of $\mathrm{V_{Si}}$ (both h and k2) have opposite phase in ESR spectra. This is a signature of the spin pumping under optical excitation, leading to the inverse population between spin sublevels. As a result,  radiofrequency (RF) emission rather than absorption is detected for one of these ESR lines \cite{SupplementalMaterial}. For  $\mathrm{V_{Si}(h)}$  the RF emission is observed at the magnetic field $B_{h-}$, as schematically indicated in Fig.~\ref{fig1}(d). For convenience we assume positive value for the absorbing resonances ($A_{h+} > 0$) and negative value for the emitting resonances ($A_{h-} < 0$). 

\begin{figure}[btp]
\includegraphics[width=.48\textwidth]{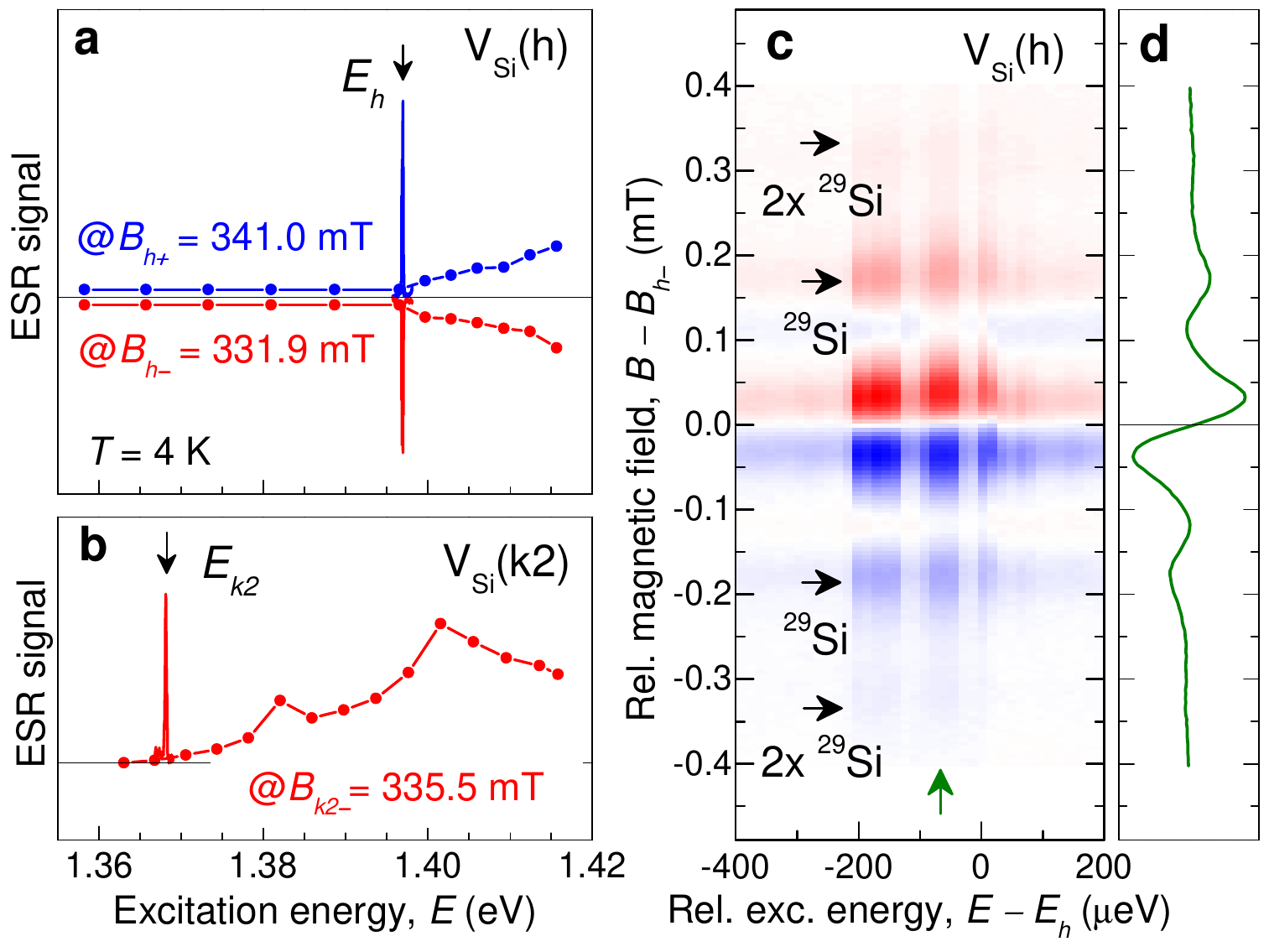}
\caption{(color online). (a) The peak-to-peak values $A_{h \pm}$  of the $\mathrm{V_{Si}(h)}$ ESR lines [see Fig.~\ref{fig1}(c)] recorded in magnetic fields $B_{h-} =331.9 \, \mathrm{mT}$ and $B_{h+} =341.0 \, \mathrm{mT}$ as a function of excitation energy.  (b) The same as (a), but for $A_{k2-}$ recorded in a magnetic field $B_{k2-} =335.5 \, \mathrm{mT}$.  (c) A 2D color plot of the ESR signal as a function of the relative magnetic field $B - B_{h-}$ and relative excitation energy $E-E_h$.  Here, $B_{h-} =331.9 \, \mathrm{mT}$  and $E_h =1.397 \, \mathrm{eV}$ correspond to the ESR and optical resonance of the $\mathrm{V_{Si}(h)}$ defect, respectively.   (d)  A cross-section of \textbf{c} at $E - E_h = - 79$~$\mathrm{\mu eV}$ (as indicated by the vertical arrow). } \label{fig2}
\end{figure}

Figure~\ref{fig2}(a) shows how $A_{h \pm}$ depend on the excitation energy. For energies $E < E_h$ the spin pumping is inefficient, for $E  > E_h$ it growths monotonically, presumably due to phonon-assisted processes and for $E = E_h$ a very sharp optical resonance is detected.  A similar behavior is observed for  $A_{k2-}$ with $E = E_{k2}$ [Fig.~\ref{fig2}(b)].  

\begin{figure}[btp]
\includegraphics[width=.47\textwidth]{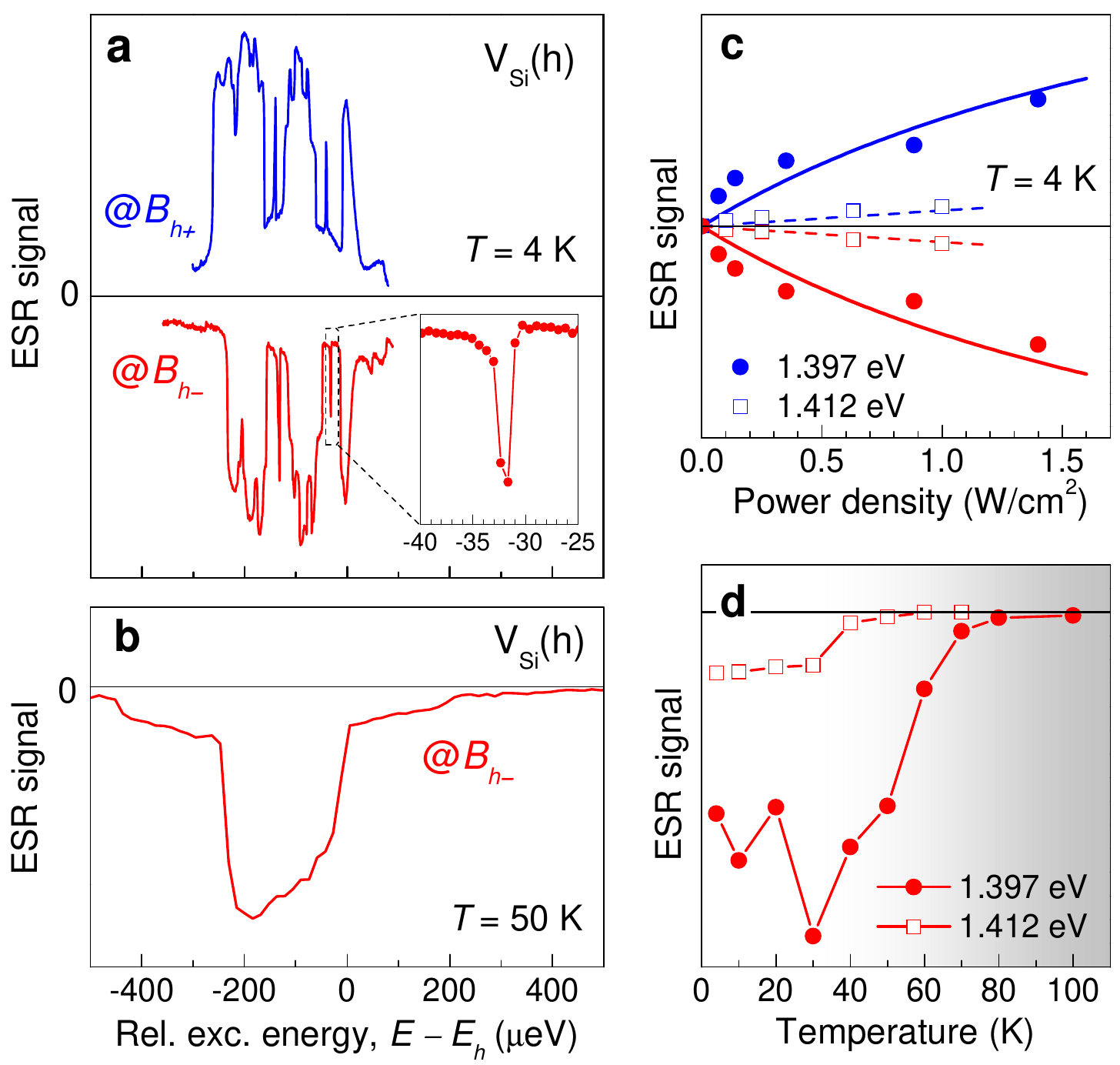}
\caption{(color online). (a,b) The same as in Fig.~\ref{fig2}(a), but recorded in the vicinity of $E_h$ with higher spectral resolution at different temperatures.  (c)  The peak-to-peak values  $A_{h \pm}$ of the $\mathrm{V_{Si}(h)}$ ESR lines recorded in magnetic fields $B_{h-}$ and $B_{h+}$ as a function of excitation power density. The solid lines are fits to Eq.~(\ref{Amplitude}). The dashed lines are linear fits. (d) The peak-to-peak values $A_{h \pm}$  under excitation into the ZPL ($E_h =1 .397 \, \mathrm{eV}$) and above-ZPL excitation ($E =1.412 \, \mathrm{eV}$) as a function of temperature.  With rising temperature the SiC sample becomes impeding the ESR detection in the grey shaded temperature region. } \label{fig3}
\end{figure}

We are now in the position to discuss the double radio-optical resonance of $\mathrm{V_{Si}(h)}$. The ESR signal as a function of the relative magnetic field $B - B_{h-}$ and relative excitation energy $E-E_h$ is presented in Fig.~\ref{fig2}(c). One can relate the ESR satellites at  $B - B_{h-} = \pm 0.17 \, \mathrm{mT}$ and  $\pm 0.34 \, \mathrm{mT}$ to one and two nuclear spin-carrying isotopes $\mathrm{^{29}Si}$ ($I = 1/2$) among the 12 next-nearest-neighbor silicon atoms [Fig.~\ref{fig2}(d)]. The optical resonance reveals a complex structure consisting of a series of extremely sharp lines. It is more clearly seen in  Fig.~\ref{fig3}(a), where the $\mathrm{V_{Si}(h)}$ optical resonance from Fig.~\ref{fig2}(a) is shown with higher resolution. Remarkably, the spectral width of an isolated line is about 2~$\mathrm{\mu eV}$ [the inset of Fig.~\ref{fig3}(a)], which comparable with a typical spectral linewidth of single defects \cite{ETuning}.  We ascribe these results to different areas within our SiC crystal, having slightly varied ZPL energies due to different local environment. 

Based on the data of Figs.~\ref{fig2} and \ref{fig3}(a) an important conclusion can be drawn: 
the  $\mathrm{V_{Si}(h)}$ spins are only addressed when the optical resonance and ESR conditions are simultaneously fulfilled. The contrast -- i.e., the ratio of the ESR signal between on and off resonant optical excitation -- is above 200.  No signature of $\mathrm{V_{Si}(h)}$ ionization is observed. Such a behavior differs from diamonds, where  for the resonant control of NV defects an additional non-resonant illumination is necessary to deshelve the NV defects from the dark state \cite{Deshelving}. To the best of our knowledge, in the solid state, similar double radio-optical resonance has only been observed so far in quantum dots \cite{QD}.  

We now discuss the efficiency of the optical resonance pumping. The $A_{h \pm}$ dependence on the excitation power density $P$ is presented in Fig.~\ref{fig3}c. In the range of $P$ under study we observe $A_{h-} \approx - A_{h+}$, meaning that the photo-induced spin polarization is significantly larger than that due to the Boltzmann statistics.  The experimental data can be well described by a standard model for spin pumping, which was initially applied for atoms \cite{SupplementalMaterial,QW}
\begin{equation}
A_{h \pm} \approx \pm A_h = \pm A_{h0} \frac{1}{1 + P_0 / P} \,.
 \label{Amplitude}
\end{equation}
Here, $P_0 = 2.62 \, \mathrm{W/cm^{2}}$ is a characteristic power density obtained in time-resolved experiments as described later.  A phonon-assisted spin pumping with the excitation energy 15~meV above the ZPL is less efficient and for $P < 1 \, \mathrm{W/cm^{2}}$ linearly depends on the laser power density. 

With rising temperature, the resonance spin pumping is observed up to 50~K [Fig.~\ref{fig3}(d)].  Remarkably, at $T = 50 \, \mathrm{K}$ the multiple-line structure is not resolved any more, but the contrast still remains the same  [Fig.~\ref{fig3}(b)]. The suppression above 50~K is accompanied  by the increase of sample conductivity due to the ionization of nitrogen donors. The conductivity is probably the only limiting factor, as in insulating SiC samples the spin pumping is efficient even at room temperature \cite{RT_SiliconVacancy}.  

\begin{figure}[btp]
\includegraphics[width=.45\textwidth]{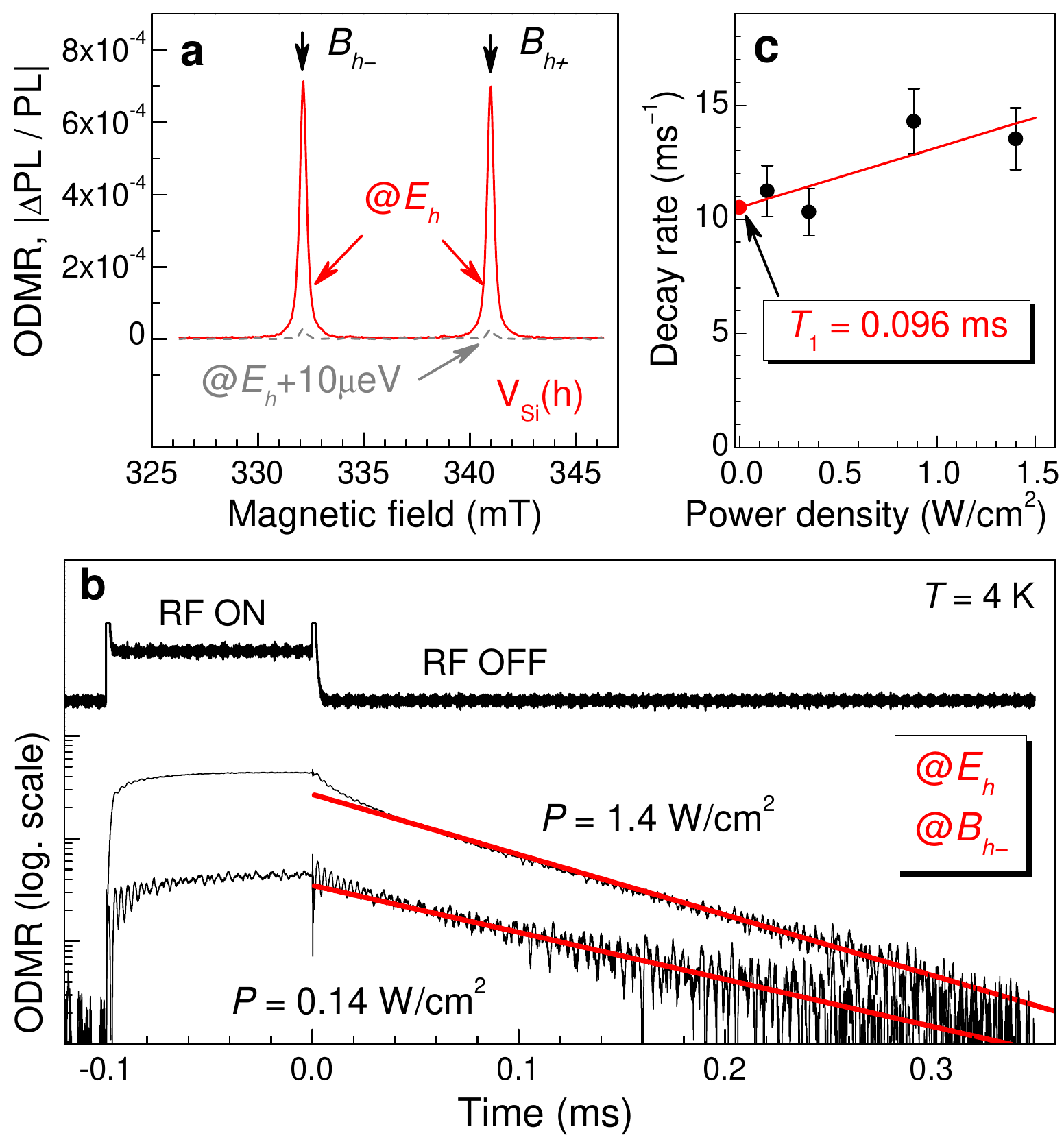}
\caption{(color online). (a) ODMR spectra, obtained under optical excitation with an energy of  the  $\mathrm{V_{Si}(h)}$ ZPL transition $E_h =1.397 \, \mathrm{eV}$ and with a slightly detuned energy $E = E_h + 10 \, \mathrm{\mu  eV}$. $B \| c$. (b) Time-resolved ODMR signal (the lower curves) recorded after a RF pulse (the upper curve) under the double spin resonance condition, i.e., with an excitation energy $E_h$ and in a magnetic field $B_{h-}$. The thick solid lines represent exponential fits $A_h \exp (-  \alpha t)$ for two excitation power densities $P = 1.4$ and $0.14 \, \mathrm{W / cm^{2}}$.  (c)  Decay rate $\alpha$ as a function of $P$.  A linear extrapolation to $P=0$ yields the spin-lattice relaxation time $T_1 = 0.096 \pm 0.011 \, \mathrm{ms}$. } \label{fig4}
\end{figure}

We now discuss possible applications of the double radio-optical resonance in SiC. Due to local environment, the optical resonance and ESR energies are individual for each defect. Additionally, they can be changed by a local electric or magnetic field in the range of  $50 \, \mathrm{\mu  eV}$ \cite{EField, BField,ETuning}. This can eventually provide a spectroscopic tool to selectively address and manipulate entangled $\mathrm{V_{Si}}$  qubits by varying the excitation energy or alternatively by tuning the double radio-optical resonance conditions for a given $\mathrm{V_{Si}}$  qubit. 

The demonstrative experiments are performed using the optically detected magnetic resonance (ODMR) technique, which can be sensitive to a single defect spin \cite{singleNV}. The ODMR signal is obtained as a normalized change in photoluminescence $\mathrm{\Delta PL / PL}$ of the sideband phonon replicas, as indicated in  Fig.~\ref{fig1}(b). We note that the $\mathrm{V_{Si}}$-related PL  is recorded even at room temperature. In the ODMR experiments the PL was passed through a 925~nm longpass filter and detected by a Si photodiode. In order to improve the signal-to-noise ratio the RF was modulated on/off at a frequency of 4.2~kHz and the photovoltage was locked-in.  

Figure~\ref{fig4}(a) shows $\mathrm{\Delta PL / PL}$ as a function of magnetic field obtained under the resonance excitation with the energy $E_h$. Two $\mathrm{V_{Si}(h)}$ lines are observed at magnetic fields $B_{h-}$ and $B_{h+}$, similar to the light-enhanced ESR experiments of Fig.~\ref{fig1}(c). When the excitation energy is detuned off the optical resonance by only 10~$\mathrm{\mu eV}$, the ODMR signal is strongly suppressed. 

Finally, we measured spin dynamics of $\mathrm{V_{Si}(h)}$ under the double radio-optical resonance conditions, i.e., at $E_h$ and $B_{h-}$. The experimental details are given in the Supplemental Material \cite{SupplementalMaterial}. In brief, the system is optically pumped with continuous wave (cw) excitation.  An intense RF pulse equalizes the spin population in the different sublevels.  The time evolution of the ODMR signal after the RF pulse follows an exponential decay. Fit examples to $A_h \exp (- \alpha t)$ for different pump power densities $P$ are presented in Fig.~\ref{fig4}b. As can be shown \cite{SupplementalMaterial},  the decay rate $\alpha$ depends on the spin-lattice relaxation time $T_1$ and characteristic power density $P_0$ as 
\begin{equation}
\alpha = \frac{1}{T_1} \left(1 + \frac{P}{P_0} \right) \,.
 \label{Time}
\end{equation}
A linear extrapolation of $\alpha (P)$ to $P=0$ yields $T_1 = 0.096 \pm 0.011 \, \mathrm{ms}$ [Fig.~\ref{fig4}(c)]. This compares reasonably well with that of NV defects in diamonds \cite{T1_diamond}. 

\textit{Conclusions and outlook.---}The silicon vacancy defects in SiC  combine the advantages of semiconductor quantum dots  and the nitrogen-vacancy defects in diamond in one system, making them very attractive for quantum spintronics applications. They are well isolated from SiC lattice, and tunable lasers with narrow linewidth can be utilized for the double radio-optical resonance coherent control of $\mathrm{V_{Si}}$ qubits, by analogy with atoms. Given that silicon and carbon have nuclear spin free isotopes, one would expect extremely long spin coherence in isotopically purified $\mathrm{^{28}Si^{12}C}$ , similar to phosphorous donors in pure $\mathrm{^{28}Si}$ \cite{Silicon28}.  An important technological aspect of $\mathrm{V_{Si}}$ in SiC is their possibility to be created in state-of-the-art transmission electron microscopes (TEMs) \cite{Steeds2002}. Considering that the  electron beam in a TEM can be focused better than 1~nm, previously unachievable perspectives for spin engineering on an atomic scale may become feasible. Furthermore, since spatially separated $\mathrm{V_{Si}}$ defects can be selectively addressed even within a single SiC nanocrystal, they can be used as a sensitive probe to image magnetic field vectors and field gradients at nanoscale. And last but not least, single $\mathrm{V_{Si}}$ centers in SiC are optically active in the NIR spectral range, overlapping with the first telecom window. This may allow for the practical realization of quantum communications as silicon vacancies can be incorporated into SiC-based LED structures or SiC-based photonic crystals. 


\begin{acknowledgments}

\textit{Acknowledgments.---}We acknowledge financial support by  the Bavarian Ministry of Economic Affairs, Infrastructure, Transport and Technology as well as by the Russian Ministry of Education and Science (contract 16.513.12.3007). 

\end{acknowledgments}

\end{document}


\maketitle

\section{Experimental methods}

\subsection{Electron spin resonance (ESR)}

\begin{figure}[htpb]
\centerline{\includegraphics[width=0.75\columnwidth,clip]{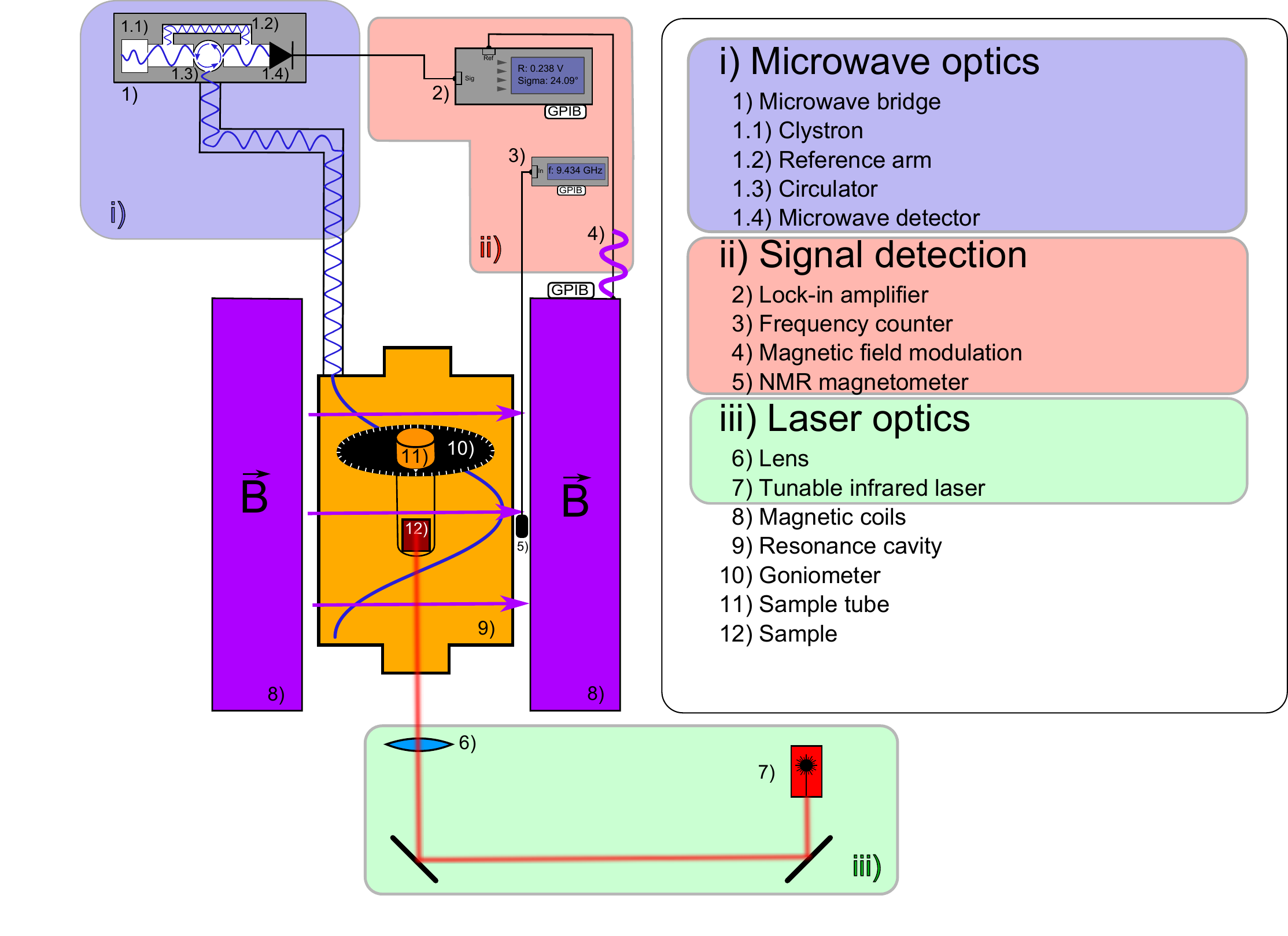}}
\caption{Sketch of the ESR setup. The investigated sample is put into a microwave resonance cavity, surrounded by magnetic coils. Through an optical access a laser beam is coupled into the system. The variation of microwave absorption is detected with a lock-in amplifier triggered by a small superimposed magnetic field. }
\label{ESR-setup}
\end{figure}

The experimental setup for performing ESR studies of the SiC samples is sketched in Fig.~\ref{ESR-setup}.  Via a quartz ESR tube, the investigated sample is placed into the center of a microwave resonance cavity surrounded by a pair of magnetic coils, which allows for applying a homogeneous external magnetic field. The cuboid cavity is the core of the home-modified X-Band spectrometer (Bruker 200D) and built in a way, that, if a microwave is coupled into it, standing waves in the transverse electric $\mathrm{TE_{102}}$ mode will emerge. The indices are related to the number of half-waves in each spatial direction. A distinctive feature of this standing wave mode is that its magnetic field maximum oscillates at the center of the cavity. The Q-factor (quality factor) of the used cavity is $Q = 2800-3200$. The microwave bridge contains the devices for generation and detection of the microwave radiation with a frequency of about $9-10 \, \mathrm{GHz}$, corresponding to $\lambda \approx 3 \, \mathrm{cm}$. An automated frequency control circuit stabilizes the adjusted resonance frequency, which is accurately read out using a frequency counter. 

For the optical excitation of the sample a tunable ($875-940 \, \mathrm{nm}$) diode laser system (Sacher Lion TEC 500) is used. The wavelength can be coarsely set with a step motor motion controller, which alters the angle of an diffraction grating built into the laser system. The laser cavity can be slightly deformed using a piezo element which allows for very fine additional tuning of the laser energy within a range of about $\pm 10 \, \mathrm{GHz}$. A special feature of the laser beam is the very narrow linewidth of about $1 \, \mathrm{MHz}$, which enables for very accurate investigation of the excitation properties. Via an optical access the sample is irradiated with the laser. The beam is widened by a lens mounted in the optical path guaranteeing illumination of the whole sample.

\begin{figure}[htpb]
\centerline{\includegraphics[width=0.6\columnwidth,clip]{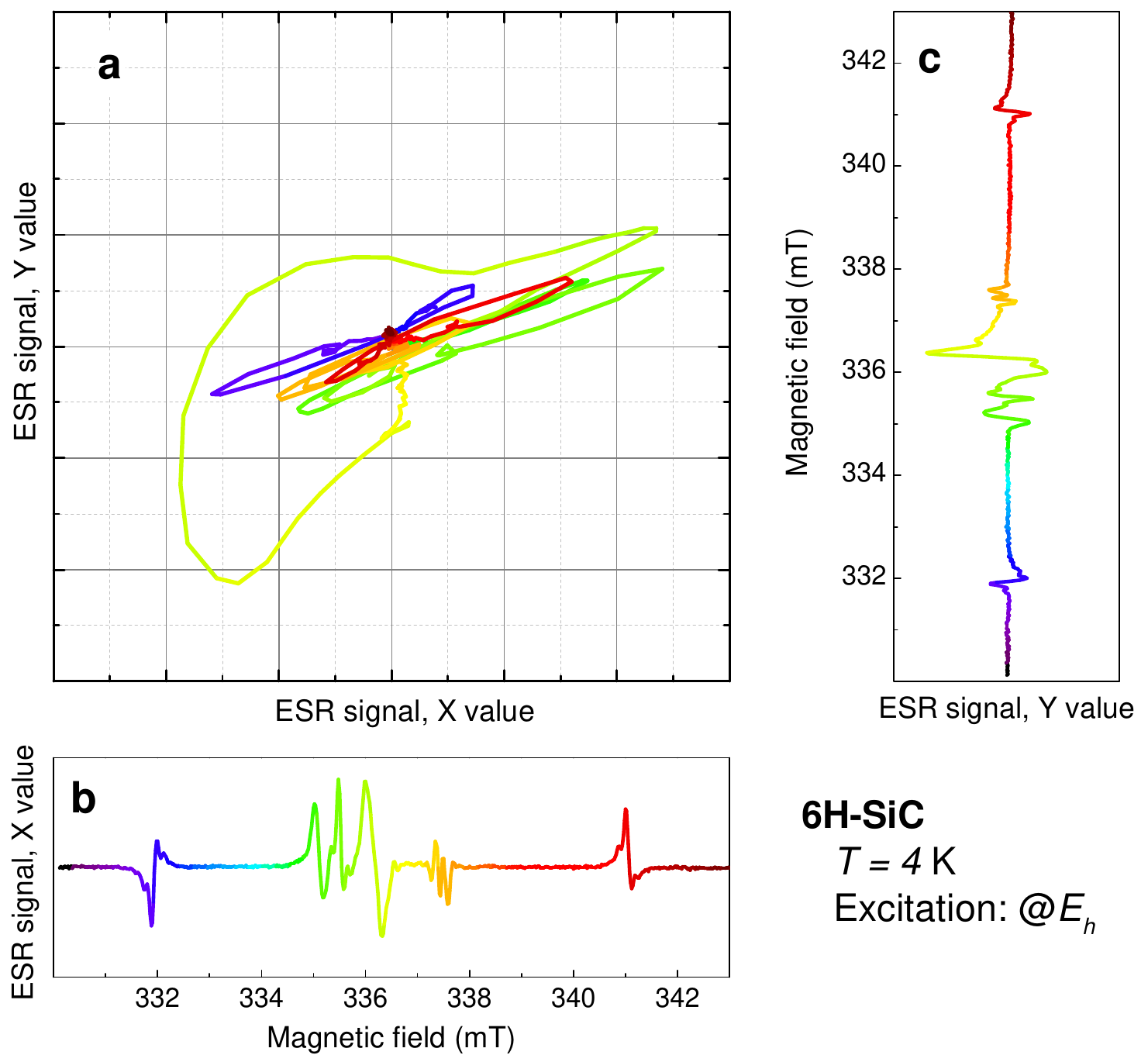}}
\caption{\textbf{a}, X-Y plot of an ESR spectrum obtained under OR excitation at the $\mathrm{V_{Si} (h)}$ ZPL of 6H-SiC. \textbf{b,c}, Corresponding X and Y values of the ESR spectrum from \textbf{a}. Magnetic field is color-coded. }
\label{ESR}
\end{figure}

In order to increase the detection sensitivity the signal obtained by the microwave detector is enhanced by a lock-in amplifier. The external magnetic field generated by the pair of coils is superimposed by a small wobble field with a frequency of $100 \, \mathrm{kHz}$. In ESR measurements, the microwave absorption is obtained while sweeping the magnetic field. Thus, the modulation results in detection of the first derivative of the signal. A certain phase shift $\theta$ of the signal occurs with respect to the reference signal of the lock-in, which is ascribed to a time delay caused by the response time of the cavity, the electronics and physical procedures. Depending on the phase setting certain amplitudes in the ESR spectrum are emphasized, while others are diminished or, in case of their phase being shifted by $90 ^{\circ}$, totally faded out. Hence, a powerful tool to distinguish and identify superimposed signals is to analyze the X-Y-plot. This can be seen in Fig.~\ref{ESR}a, where the phase diagram of an ESR spectrum is presented.  The magnetic field in the figure is coded by color, and the signal phases at 332.0 and 341.0~mT are obviously different from that at 336.0~mT. By variation of the detecting phase it is possible to highlight different resonances (Figs.~\ref{ESR}b and c). We use this technique to maximize the signal amplitude of the outer ESR lines ascribed to the $\mathrm{V_{Si} (h)}$ defect. 

To improve the accuracy of the detection of the external magnetic field a NMR magnetometer is mounted next to the cavity. The Larmor frequency of protons in a reference material with known g-factor is measured and read out by a frequency counter.

\subsection{Optically detected magnetic resonance (ODMR)}

\begin{figure}[htpb]
\centerline{\includegraphics[width=0.75\columnwidth,clip]{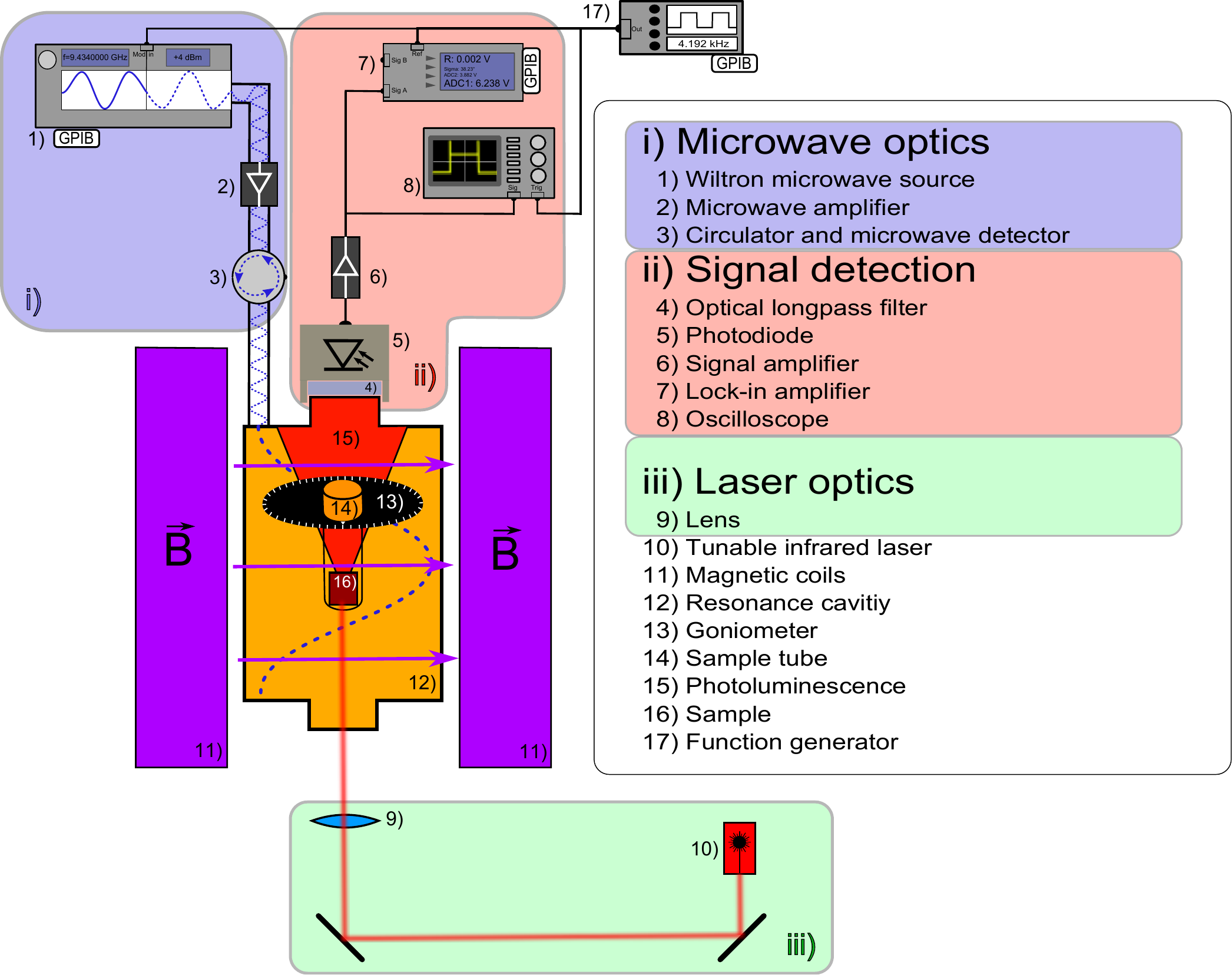}}
\caption{Scheme of the ODMR setup. The investigated sample is put into a microwave resonance cavity surrounded by magnetic coils. Via optical accesses the sample is illuminated with an infrared laser and the emitted light is obtained with a silicon detector. Sweeping the magnetic field the relative change in PL intensity is obtained for a chopped microwave irradiation using lock-in detection. The transient signals are recorded using an osciloscope.}
\label{ODMR-setup}
\end{figure}

The ODMR setup shown in Fig.~\ref{ODMR-setup} is a variation of the ESR setup presented in the previous section. The cavity is designed for optical access from the front and the rear side allowing for coupling of a laser beam and optical detection. The photoluminescence (PL) is detected in a transmission geometry by a Si photodiode and the laser emission is filtered out by a 925~nm longpass filter. A different microwave source is used (Wiltron 69137A Sweep Generator), which features the variation of the emitted microwave power and the possibility to chop the microwave by an external driver. The maximum accessible output power of the microwave source is $60 \, \mathrm{mW}$, which is additionally amplified reaching a maximum power within the cavity of $2 \, \mathrm{W}$. The change in PL intensity induced by the change in the ground state population is locked-in. For taking into account that the overall PL is governed by different external factors and for enabling to compare different spectra the change in PL is normalized, $\mathrm{\Delta PL / PL}$ . 

In transient ODMR experiments, instead of using a lock-in amplifier the pre-amplified signal is detected with an oscilloscope (LeCroy WaveRunner 610zi) triggered by the chopped microwave source. 
In order to improve the signal-to-noise ratio the signal is averaged over 1000 scans. Using a function generator the pulse length of the microwave can be varied in order to systematically investigate the population difference of the involved spin levels, which is proportional to the obtained ODMR signal due to spin-selective PL.

\section{Optical spin pumping of the silicon vacancy}

\subsection{Absorption and emission of the silicon vacancy}

\begin{figure}[htpb]
\centerline{\includegraphics[width=0.85\columnwidth,clip]{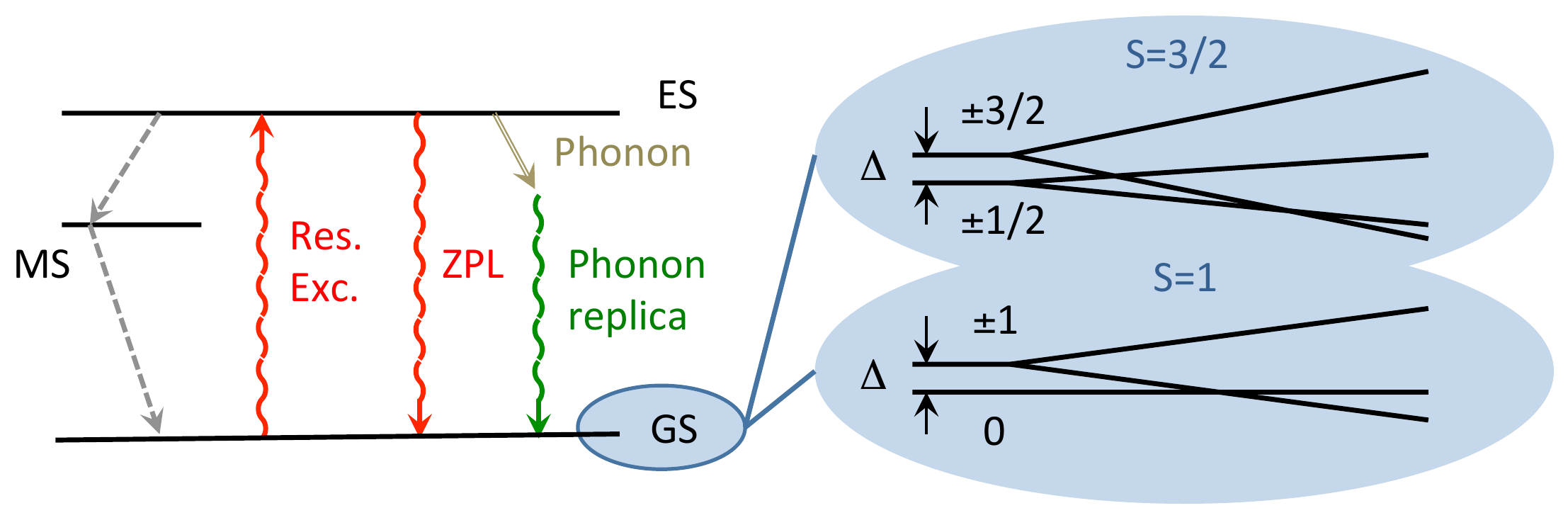}}
\caption{Scheme of the resonance excitation (Res. Exc.) and zero-phonon line (ZPL) emission between the ground state (GS) and the excited state (ES). Radiative recombination accompanied by a photon emission results in phonon replica. The GS has high-spin state with spin  multiplicity either $S = 1$ or $S = 3/2$. Spin polarization under optical excitation occurs via spin dependent relaxation through the metastable state (MS). }
\label{Pump}
\end{figure}

Considering a defect center embedded into a crystal, the interaction with light can be explained by the interplay of electronic and vibronic transitions between different energy levels (Fig.~\ref{Pump}). The transition between the lowest vibrational levels of the ground state (GS) and the excited state (ES) not involving phonons is called zero-phonon line (ZPL). The phonon transitions of all lattice modes contribute and form a continuous phonon replica band of the zero phonon line, which is blue-shifted for absorption respectively red-shifted for PL. While a sharp ZPL can be observed at cryogenic temperatures, the probability for phonon-assisted transitions increases with temperature due to thermal activation. 

One of the special properties of the investigated silicon vacancy defects constitutes the possibility of polarizing their ground state. The working model of the energy structure and optical pumping scheme can be adopted from that for NV defects in diamond. The photo-kinetic processes leading to spin alignment under optical pumping can be described by the  presence of an excited metastable state (MS) and a spin-dependent inter-system crossing (ICS) between the high-spin ES, the MS and the high-spin GS (Fig.~\ref{Pump}). The optical transitions between the GS and the ES are spin-conserving. Remarkably, the nonradiative decay via the MS is spin-selective resulting in a polarization of the GS. Depending on the polytype of SiC, the lattice site and temperature the transition rates between the energy levels vary resulting in different population of the spin sublevels.

\subsection{ESR and ODMR spectra of the silicon vacancy}

The spin Hamiltonian of the silicon vacancy in magnetic field can be written in the form 
\begin{equation}
\mathcal{H} = g_e \mu_B  \mathbf{S} \mathbf{B} + D [ S_z^2 - \frac{1}{3} S(S+1) ] \,.
 \label{Hamilton}
\end{equation}
Here, $S_z$ is the spin projection on the $c$ axis of SiC crystal. The parameter $D$ characterizes the axial crystal field.  There is still no agreement in the community whether the ground state of $\mathrm{V_{Si}}$ in SiC has spin $S = 1$ or $S = 3/2$. We now show that the ESR and ODMR spectra presented in this work are indistinguishable between these cases. 

\begin{figure}[htpb]
\centerline{\includegraphics[width=0.9\columnwidth,clip]{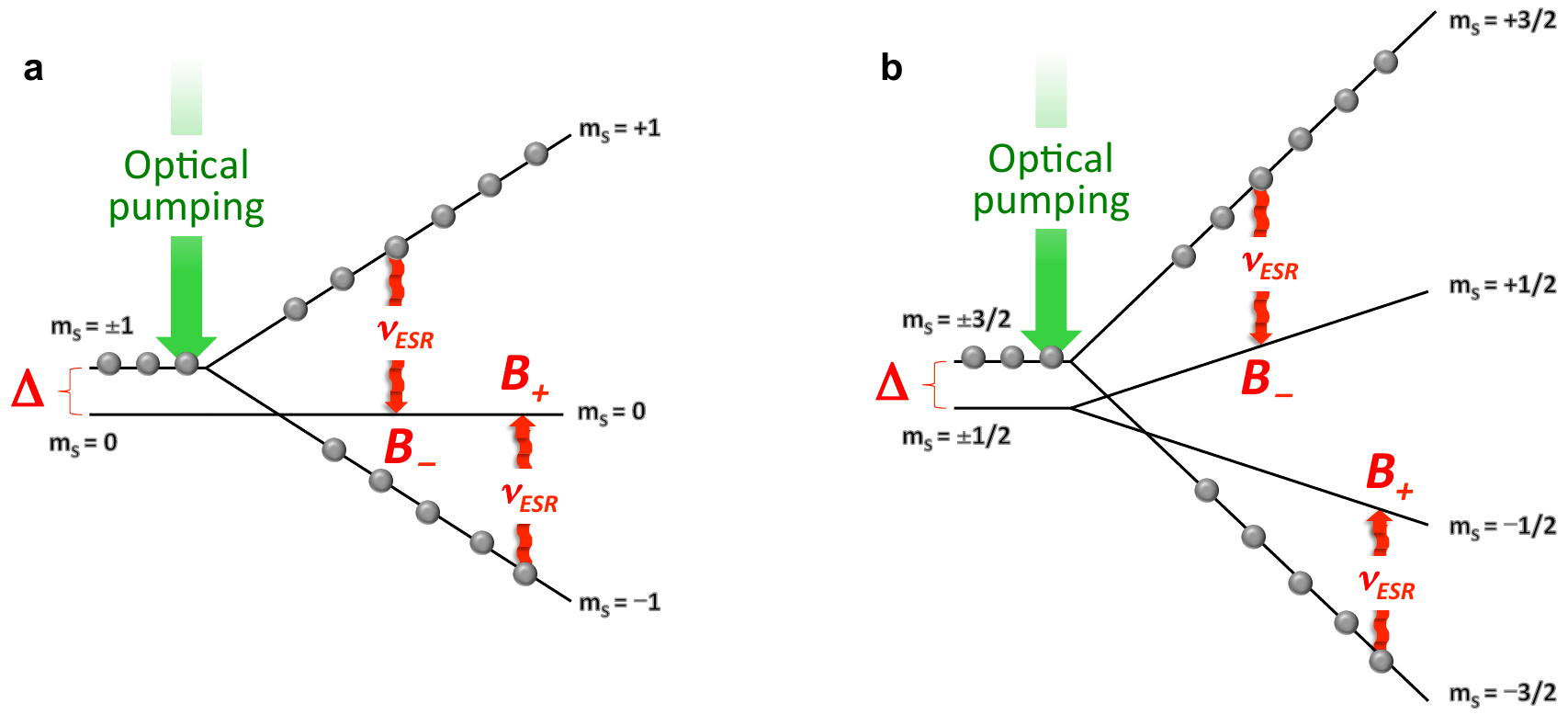}}
\caption{\textbf{a}, A scheme of the light-enhanced ESR in case when the ground state has spin $S = 1$. \textbf{b}, A scheme of the light-enhanced ESR in case when the ground state has spin $S = 3/2$. }
\label{SpinState}
\end{figure}

\subsubsection{\textit{S} = 1 ground state}

Following Eq.~(\ref{Hamilton}),  zero field splitting $\Delta = D$ occurs between the $m_s = 0$ and $m_s = \pm 1$ spin sublevels (Fig.~\ref{SpinState}a). In an external magnetic field $B$ applied parallel to the $c$ axis of SiC crystal the $m_s = \pm 1$ sublevel is split further as $E_{\pm1} = \pm g_e \mu_B B$.  In the first order, the allowed transitions are with $\delta m_s = \pm 1$ and the ESR for the transitions $+1 \rightarrow 0$ and $-1 \rightarrow 0$ occur at different magnetic fields, labeled in Fig.~\ref{SpinState}a as $B_-$ and $B_+$ respectively. There is the following interconnection between these fields and $\Delta$ 
\begin{equation}
h \nu_{ESR}= \pm \Delta + g_e \mu_B B_{\mp} \,.
 \label{ZFS1}
\end{equation}
Here, $\nu_{ESR}$ is the resonance frequency of the ESR spectrometer. Under illumination the system is pumped into the $m_s = \pm 1$ states. This implies inverse population: at $B_+$ the RF radiation is absorbed, while at $B_-$ the RF
radiation is emitted. This is what we observe in our light-enhanced ESR experiments (see Figs.~1c and d of the main text).  

Remarkably, the average photoemission rate involving the $m_s = 0$ and $m_s = \pm 1$ spin sublevels is different. Resonant RF changes spin population, resulting in a change of the PL intensity ($\mathrm{\Delta PL / PL}$). The corresponding ODMR spectrum should reveal two lines, as we observed in our experiments (see Fig.~4a).  

\subsubsection{\textit{S} = 3/2 ground state}

Following Eq.~(\ref{Hamilton}), zero field splitting $\Delta = 2 D$ occurs between the $m_s = \pm 1/2$ and $m_s = \pm 3/2$ spin sublevels (Fig.~\ref{SpinState}b). In an external magnetic field $B$ applied parallel to the $c$ axis of SiC crystal the $m_s = \pm 1/2$ and $m_s = \pm 3/2$ sublevels are split further as $E_{\pm1/2} = \pm g_e \mu_B B / 2$ and $E_{\pm3/2} = \pm 3 g_e \mu_B B / 2$ respectively.  In the first order, the allowed transitions are with $\delta m_s = \pm 1$ and the ESR for the transitions $+3/2 \rightarrow +1/2$ and $-3/2 \rightarrow -1/2$ occur at different magnetic fields, labeled in Fig.~\ref{SpinState}a as $B_-$ and $B_+$ respectively. There is the following interconnection between these fields and $\Delta$ 
\begin{equation}
h \nu_{ESR}= \pm \Delta + g_e \mu_B B_{\mp} \,,
 \label{ZFS32}
\end{equation}
which is equal to Eq.~(\ref{ZFS1}). Under illumination the system is pumped into the $m_s = \pm 3/2$ states. Again, this implies inverse population: at $B_+$ the RF radiation is absorbed, while at $B_-$ the RF radiation is emitted. The third allowed transition  $-1/2 \rightarrow +1/2$ should appear at the magnetic field $B_0 = (B_+ + B_- ) / 2$. However, it is insensitive to the illumination and indistinguishable from other resonances appearing in ESR spectra at $B_0$ (see Figs.~1c and d). 

The average photoemission rate involving the $m_s = \pm 1/2$ and $m_s = \pm 3/2$ spin sublevels is different. Therefore, redistribution in the spin population between the $m_s = - 1/2$ and $m_s = + 1/2$ sublevels does not result in a change of the PL intensity and the corresponding ODMR spectrum should again reveal two lines (see Fig.~4a).

\subsection{Model of the silicon vacancy resonant spin pumping}

We consider a two-level system with spin populations denoted as $n_+$ and $n_-$. In case of the $S = 1$ ground state, $n_+$ corresponds to the aggregate population of the $m_s = \pm 1$ sublevels and  $n_-$ corresponds to the population of the $m_s = 0$ sublevel. In case of the $S = 3/2$ ground state, $n_+$ corresponds to the aggregate population of the $m_s = \pm 3/2$ sublevels and  $n_-$ corresponds to the aggregate population of the $m_s = \pm 1/2$ sublevels. 

An intense RF pulse equalizes the spin population and at time $t = 0$ we have $n_+ (0) = n_- (0)= \frac{1}{2}$. After the RF pulse the system is optically pumped, i.e.,  the population $n_+$ increases and the population $n_-$ decreases. The spin pumping rate is higher the higher the laser power density $P$ and the higher the population $n_-$ are.  Simultaneously spin relaxation with the spin-lattice relaxation time $T_1$ occurs. It is proportional to the difference from the population in the thermal equilibrium $n_{eq \pm}$, which can be well approximated to $n_{eq-} \approx n_{eq+} \approx \frac{1}{2}$. Finally, the following differential equations can be written
\begin{eqnarray}
\frac{d}{dt} n_+  =   k P n_-  - \frac{1}{T_1}  \left( n_+ - \frac{1}{2}  \right) \,,     \label{Pump1a}  \\
\frac{d}{dt} n_-  =   -k P n_-  - \frac{1}{T_1}  \left( n_- - \frac{1}{2}  \right) \,.     \label{Pump1b}
\end{eqnarray}
Here, $k$ is a coefficient of proportionality, such that $k P n_-$ defines the spin pumping rate. The ESR signal is proportional to the population difference $\Delta n = n_+ - n_-$, and with $n_+ (t) + n_- (t) = 1$ we obtain
\begin{equation}
\frac{d}{dt} \Delta n  =  k P - \- \left( k P  + \frac{1}{T_1}  \right) \Delta n \,. 
\label{Pump3}
\end{equation}
With boundary condition $\Delta n (0) = 0$ the solution of Eq.~(\ref{Pump3}) is 
\begin{equation}
\Delta n  (t) =  \Delta n_\infty  \- \left( 1 - e^{- \alpha t} \right)
\label{Pump4}
\end{equation}
with spin relaxation rate
\begin{equation}
\alpha =  \frac{1}{T_1}  \left( 1 + \frac{P}{P_0} \right)
\label{Pump5}
\end{equation}
and steady-state value
\begin{equation}
\Delta n_\infty =  \frac{1}{1 + P_0 / P}  \,. 
\label{Pump6}
\end{equation}
Here, $P_0 = \frac{1}{k T_1}$ is a characteristic power density.